\documentclass[12pt,reqno]{amsart}

\usepackage[english]{babel}
\usepackage{mathbbol}
\usepackage{amsmath}
\usepackage{amsthm}
\usepackage{amsfonts}
\usepackage{amssymb}
\usepackage[citation-order,short-journals]{amsrefs}

\BibSpec{article}{%
  +{}{\PrintAuthors}{author}
  +{,}{ \textit}{title}
  +{,}{ }{journal}
  +{}{ \textbf}{volume}
  +{,}{ }{pages}
  +{}{ \parenthesize}{date}
  +{,}{ \texttt}{eprint}
  +{.}{}{transition}
}

\allowdisplaybreaks[3]
 
\newcommand{\N}{\mathbb{N}}
\newcommand{\Z}{\mathbb{Z}}
\newcommand{\R}{\mathbb{R}}
\newcommand{\C}{\mathbb{C}}

\newcommand{\I}{\Eins}
\newcommand{\cst}{\ensuremath{\mathcal{C}^\ast}}
\newcommand{\A}{\ensuremath{\mathfrak{A}}}
\newcommand{\Lat}{\ensuremath{\mathfrak{L}}}
\newcommand{\Hi}{{\mathcal{H}_J}}
\newcommand{\ccr}{\mathsf{CCR}}
\newcommand{\hlf}[1][1]{\scriptscriptstyle{#1/2}}

\newcommand{\underbr}[2]{\ensuremath{\underset{#1}{\underbrace{#2}}}}
\newcommand{\defeq}{:=}

\theoremstyle{plain}
\newtheorem{theorem}{Theorem}[section]

\newtheorem{proposition}[theorem]{Proposition}
\newtheorem{lemma}[theorem]{Lemma}
\newtheorem{corollary}[theorem]{Corollary}
\newtheorem*{theorem*}{Theorem}
\theoremstyle{definition}
\newtheorem{remark}[theorem]{Remark}

\DeclareMathOperator*{\slim}{s-lim}
\DeclareMathOperator*{\otim}{\otimes}

\renewcommand{\phi}{\varphi}
\renewcommand{\epsilon}{\varepsilon}
\renewcommand{\Im}{\mathfrak{Im}\;}

\begin{document}

\title[Central limit theorems for the asymptotics of quantum spins]{Central 
limit theorems for the large-spin asymptotics of quantum spins}

\author{Tom Michoel}
\address{Instituut voor Theoretische Fysica,
Katholieke Universiteit Leuven,
Celestijnenlaan 200 D,
B--3001 Leuven,  Belgium}
\email{tom.michoel@fys.kuleuven.ac.be}
\author{Bruno Nachtergaele}
\address{Department of Mathematics, University of California, Davis, 
One Shields Avenue, Davis, CA 95616-8366, USA}
\email{bxn@math.ucdavis.edu}
\thanks{This material is based on work supported by the National Science
Foundation under Grant No.~DMS0303316. T. Michoel
is a Postdoctoral Fellow of the Fund for
Scientific Research -- Flanders (Belgium) (F.W.O.--Vlaanderen)}
\thanks{\copyright\ 2003 by the authors. This article may be reproduced in its
entirety for non-commercial purposes.}


\date{10 October 2003}


\begin{abstract} We use a generalized form of Dyson's spin wave formalism to
prove several central limit theorems for the large-spin asymptotics of quantum
spins in a coherent state. \end{abstract}

\maketitle

\section{Introduction}
\label{sec:intro}

In statistical mechanics, the thermodynamic limit of an infinite number of interacting
particles in the continuum or on a lattice can be taken rigorously using the laws of
probability. In a first approximation we are usually interested in the behavior of
intensive observables, i.e., observables that grow proportionally to the total number of
particles. Taking their thermodynamic limit corresponds to the Law of Large Numbers
(\textsf{LLN}) in probability theory. Introducing quantum mechanics at the microscopic
level does not change much at the macroscopic level: intensive observables still behave
classically in the thermodynamic limit, or, the \textsf{LLN} for quantum systems, in
particular, involves a description of the system in terms of classical, i.e., commuting,
variables.

The next logical step is to study fluctuations of intensive observables around their mean.
The corresponding law in probability theory is the Central Limit Theorem (\textsf{CLT}).
Here, quantum effects can survive in the limit of an infinite number of particles, and
fluctuations often behave non-classically and have to be modeled using non-commuting
variables. Typically, this happens in the presence of a spontaneously broken continuous
symmetry, and the macroscopic quantum fluctuations present in such a system are the
well-known Goldstone bosons \cite{anderson:1984}. A general theory of Goldstone bosons
using non-commutative central limit theorems was presented in \cite{michoel:2000}.

The study of non-commutative central limit theorems becomes essential in this context, and
a general theory was developed in \cites{goderis:1989b, goderis:1990}. The main result is
that the macroscopic quantum fluctuations can be identified with a representation of the
Canonical Commutation Relations (\textsf{CCR}) in a quasi-free state which is determined
by the correlations in the microscopic state. The non-commutative central limit theorem
can be considered as the first quantum correction to the \textsf{LLN} for these systems.

In addition to the thermodynamic limit (i.e., the large-$N$ limit), also the so-called
classical limit results in an asymtotic description of the system by classical random
variables. A well-known example is the classical limit of quantum spins, which was treated
rigorously in great detail by Lieb \cite{lieb:1973}. This is the limit of infinite spin:
the quantum spin operators are normalized such that, in a suitable sense, they converge to
classical spin variables with values in the unit sphere in $\R^3$.

As an example of such a result we mention the classical limit of the free energy, obtained
for a large class of models by Lieb \cite{lieb:1973}. In this case, the classical limit
result tells us that after rescaling each spin operator in the Hamiltonian, the partition
function of the quantum model will converge to the partition function of the corresponding
classical model. One may also consider the limit of large spin in a sequence of states,
and compute expectations, in which case one is interested in the classical limit as a
result about the distribution of the rescaled spin operators considered as random
variables. In this present work we are interested in the latter situation.

In Lieb's treatment of the classical limit an important role is played by the so-called
coherent states, which were studied in great detail in \cite{arecchi:1972}. Among all
quantum states, the coherent states are those that optimally approximate the idea of a
spin pointing in a certain direction in space. Mathematically this is made precise by
showing that these states carry no dispersion for the classically rescaled spin operators
(see Proposition \ref{pro:cl-lim}). Again, we see that the classical limit for spins,
which is similar to a \textsf{LLN} in probability, results in classical random variables.
The role of $N$ is played by the magnitude of the spin, which we will denote by $J$. It is
now reasonable to expect that in such states in which the `intensive' (scaled) observables
become dispersionless in the limit, that a central limit theorem should hold for the
fluctuations of these observables around their expected mean. The subject of this paper is
then the formulation and proof of various central limit theorems for these fluctuations,
which are represented by operators on a Hilbert space. The class of states we consider are
products of so-called coherent states for the spins.

In Section \ref{sec:clt-gvv}, which also contains the mathematical setup including the
definition and basic properties of the coherent states, we formulate a first
non-commutative Central Limit Theorem (\textsf{CLT}) for large $J$ using the techniques of
Goderis, Verbeure, and Vets \cites{goderis:1989b, goderis:1990}.  In a rather different
setup, a similar result was previously given by Michoel and Verbeure in
\cite{michoel:1999c}. The proof of this \textsf{CLT} and some auxiliary results are given
in Appendix \ref{sec:proof-central-limit-intro}.

The main purpose of this paper is to strengthen the results of Section \ref{sec:clt-gvv},
which could be called standard, by taking advantage of the additional structure that is
present in models with a high dimensional representation of the $SU(2)$ commutation
relations.  More precisely, we want to use Dyson's spin wave formalism
\cites{dyson:1956,dyson:1956b}, which can be briefly described as follows. The `all spin
up' state is used as a reference state in spin Hilbert space. It plays the role of the
vacuum state in the sense that there is a formal analogy between lowering the spin in this
state, and creating a boson particle in a Fock vacuum state. Dyson made this analogy
precise by defining a unitary equivalence between the spin Hilbert space and a subspace of
Fock space (the subspace with no more particles present than the size of the spin). Under
this equivalence, the spin lowering operator, divided by the square root of the size of
the spin, becomes the boson creation operator up to a correction that is a function of the
number operator that formally goes to one in the large-spin limit. After the trivial
observation that every coherent state is the `all spin up' state for the spin in the
defining direction of the coherent state, we see that Dyson's formalism can be used to
study the coherent states of our situation. The spin wave formalism is discussed in detail
in Section \ref{sec:dyson}.

The scaling with the square root of the spin in Dyson's equivalence between the spin
lowering and the boson creation operator is precisely the scaling we need in the central
limit theorem for the fluctuations of the spin operators. Hence, for all values of the
spin, we can write the fluctuation operators as well-defined operators on one and the same
Fock space, and it follows (see Section \ref{sec:clt-spin-wave}), that the limit of
infinite spin can be taken as a genuine operator limit. This is where the central limit
theorem for spin fluctuations differs fundamentally from the known non-commutative central
limit theorems and it makes it possible to prove stronger results. For instance, the usual
central limit theorems prove convergence of the characteristic function of linear
combinations of fluctuation operators. Using Dyson's spin wave formalism, we obtain
convergence of the characteristic function of arbitrary polynomials of fluctuation
operators.

In the last section, Section \ref{sec:applications}, we discuss some applications of this
stronger version of the central limit theorem.

The first application (in Section \ref{sec:bosonization}) is to obtain a rigorous version
of ``bosonization'' for quantum spin systems. This refers to the well-known technique in
physics that the low-energy excitations, and hence the low-temperature behavior, of
certain quantum spin systems can be well approximated using a boson or spin wave
approximation.  This idea can be made mathematically precise for those quantum spin
systems which possess a coherent ground state. Indeed, using our central limit theorem we
immediately obtain strong convergence of the spin Hamiltonian and the spin dynamics to a
quasi-free boson Hamiltonian and dynamics. In \cite{michoel:2003a}, we use this
convergence to obtain the large-spin asymptotics of the energy spectrum of the anisotropic
ferromagnetic Heisenberg chain, thus improving upon earlier results \cite{koma:2001,
  caputo:2003}. We also discuss another application of this convergence result, namely to
obtain the large-spin asymptotics for the time evolution of the ground state when the
dynamics is perturbed by a spin fluctuation.

In a second application (Section \ref{sec:strong-clt-N}), we apply the theorem to the
study of non-commuting fluctuation operators for $N$ independent copies of quantum random
variables. A central limit theorem for the characteristic function of polynomials of such
fluctuation operators, analogous to the one we obtain in Section \ref{sec:clt-spin-wave},
has appeared in the literature in the form of a conjecture by Kuperberg
\cite{kuperberg:2002}. The special case of this conjecture for expectations computed in a
tracial state, was proved in \cite{kuperberg:2002b} and establishes an interesting result
about the distribution of the shape of a random word. This is an example of a quantum CLT
used to prove results in classical probability theory, in this case generalizing earlier
work of Johansson \cite{johansson:2001}.

The situation with a tracial state can be regarded as intermediate between classical and
quantum probability theory, because the cyclicity of the trace implies that although the
microscopic variables do not commute, the limiting fluctuations are classical Gaussian
random variables. We can apply our results to obtain the first example of a non-tracial
state for which Kuperberg's conjecture holds. We show that a system of $N$ independent
spin-$\frac12$ particles, each in a coherent state, can be identified with one spin-$\frac
N2$ particle in the corresponding coherent state. Therefore, our general results can be
applied directly and we obtain a fully non-commutative system in which the conjecture is
valid.

\section{Mathematical setup and preliminary results}
\label{sec:clt-gvv}

We consider quantum spin systems on a finite or infinite lattice $\Lat$. At each site
$x\in\Lat$ we have a spin-$J$ degree of freedom ($J\in\frac12 \N_0$), i.e., a
($2J+1$)-dimensional irreducible representation of $SU(2)$, and we denote with $S^i_x$ the
corresponding spin-$J$ matrices,
\begin{align*}
  [S^i_x,S^j_y]&=i\delta_{x,y}\epsilon_{ijk} S^k_x\\
  S_x\cdot S_x&= (S_x^1)^2 + (S_x^2)^2 + (S_x^3)^2 = J(J+1)
\end{align*}
We will also use the spin raising and lowering operators: $S^+_x$ and $S^-_x$,
$S^\pm_x=S^1_x\pm i S^2_x$.

The local Hilbert spaces associated to each finite subset $\Lambda$ of the lattice are
therefore
\begin{equation*}
  \mathcal{H}_{J,\Lambda} = \bigotimes_{x\in\Lambda} \Bigl( \C^{2J+1}\Bigr)_x
\end{equation*}
and the algebra of observables is
\begin{equation*}
  \A_{J,\Lambda} = \bigotimes_{x\in\Lambda} \Bigl( \mathbb{M}_{2J+1}(\C)\Bigr)_x
\end{equation*}
For $\Lambda^\prime\subset\Lambda$, $\A_{J,\Lambda^\prime}$ can be considered
as a subalgebra of $\A_{J,\Lambda}$ in a natural way. With this in mind,
the union
\begin{equation*}
  \A_{J,\text{loc}} = \bigcup_{\Lambda\subset\Lat} \A_{\Lambda,J}
\end{equation*}
is the algebra of local observables, and its closure is the quasi-local algebra of
observables $\A_J$ of the spin system on $\Lat$. For a finite lattice $\Lat$ this is of
course just the local algebra specified before.

As a basis for the local Hilbert spaces we take the standard basis which is diagonal for
the $S^3_x$ operators:
\begin{align*}
  S^3_x |m_x\rangle &= m_x |m_x\rangle,\quad m_x=-J, -J+1,\dots,J\\
  S^\pm_x |m_x\rangle &= \sqrt{J(J+1)-m_x(m_x\pm 1)} |m_x\pm 1\rangle
\end{align*}

A coherent spin state \cites{arecchi:1972,lieb:1973} at site $x$ is specified by a unit
vector in $\R^3$ (i.e., a \emph{classical} spin) $u_x=(\theta_x,\phi_x)$,
$0\leq\theta_x\leq\pi$, $0\leq \phi_x\leq 2\pi$, and defined by
\begin{align*}
  |(\theta_x,\phi_x)\rangle &= e^{\frac12\theta_x( S^-_x e^{i\phi_x} -
      S^+_x e^{-i\phi_x})}|J\rangle\\
  &= \sum_{m_x=-J}^J \binom{2J}{J-m_x}^{1/2}(\cos\tfrac12\theta_x)^{J+m_x}
  (\sin\tfrac12\theta_x)^{J-m_x} e^{i(J-m_x)\phi_x}\;|m_x\rangle
\end{align*}

The states of the spin system that we will consider are tensor products of coherent
states:
\begin{align*}
  \omega (A) = \otim_{x\in\Lat} 
  \langle(\theta_x,\phi_x)| A |(\theta_x,\phi_x)\rangle
\end{align*}
which is well-defined for all $A\in\A_J$. The collection of unit vectors
$\{u_x=(\theta_x,\phi_x)\}_{x\in\Lat}$ defining the state is completely arbitrary. The
\textsf{GNS} Hilbert space of $\omega$ is the incomplete tensor product Hilbert space
\begin{equation*}
  \Hi = \overline{ \bigcup_{\Lambda\subset\Lat} 
  \Bigl( \bigl[ \otim_{x\in\Lambda}
    \C^{2J+1} \bigr] 
    \otimes \bigl[ \otim_{y\in\Lat\setminus\Lambda} \Omega_y\bigr] \Bigr)}
\end{equation*}
where $\Omega_y$ is short-hand for $\Omega_y = |(\theta_y,\phi_y)\rangle$, and
the generating vector is
\begin{equation*}
  \Omega_J = \otim_{x\in\Lat} \Omega_x
\end{equation*}
Because of the simplicity of the \textsf{GNS} representation we will not need to
distinguish between an observable and its representative in $\mathcal{B}(\mathcal{H}_J)$.

A particular type of quasi-local observables that we are interested in are those of the
form
\begin{equation}\label{eq:1}
  \sum_{x\in\Lat} v_x\cdot S_x,\quad v_x\in\R^3
\end{equation}
In order that functions, such as polynomials and trigonometric functions, of these
observables are well-defined, we need summability conditions on the set of $v_x$. Consider
$v=\{v_x\in\R^d\}_{x\in\Lat}$, and define the $p$-norm ($p\geq 1$) by
\begin{equation*}
  |v|_p \defeq \Bigl( \sum_{x\in\Lat} |v_x|^p\Bigr)^{\scriptstyle{1/p}}
\end{equation*}
where $|v_x|$ is the length of $v_x$.  The space of all $\R^d$-valued summable sequences
(i.e., $|v|_1<\infty$) on $\Lat$ will be denoted $\ell^1_d(\Lat)$, and the space of all
$\R^d$-valued square summable sequences will be denoted $\ell^2_d(\Lat)$. For $d=2$ we get
the standard complex valued (square) summable sequences, and for these spaces the
subscript $d$ is omitted. Note that if $|v|_{p_0}<\infty$ for some $p_0\geq 1$, then for
$p\geq p_0$, $|v|_p\leq |v|_{p_0}<\infty$ as well.

In \cites{arecchi:1972,lieb:1973}, a number of generating functions are derived for
coherent states. The most important one for us is the characteristic function for
observables of the type \eqref{eq:1}:
\begin{equation}\label{eq:2}
  \omega\bigl( e^{i\sum_x v_x\cdot S_x}\bigr) = \prod_{x\in\Lat} 
  \Bigl\{ \cos(\frac12
  |v_x|)+ i \frac{v_x\cdot u_x}{|v_x|} \sin (\frac12 |v_x|) \Bigr\}^{2J}
\end{equation}
It follows immediately that
\begin{equation*}
  \omega\bigl( \sum_x v_x\cdot S_x\bigr) = J \sum_x v_x\cdot u_x
\end{equation*}

The classical limit of a quantum spin system is obtained by normalizing each spin operator
by $J$, and then taking the limit $J$ to $\infty$. The quantum spin variables then
converge to classical spins taking values in the unit sphere in $\mathbb{R}^3$. There are
various precise mathematical statements that can express this. The most common are results
formulated as convergence of the free energy of a quantum spin system to the free energy
of the corresponding classical spin system \cite{lieb:1973}. In the present paper, we don
not specify a Hamiltonian, but consider the characteristic functions of the spin variables
in states that are products of coherent states. The following result, proved in Appendix
\ref{sec:proof-central-limit-intro} as an easy consequence of eq. \eqref{eq:2}, is a law
of large numbers for the quantum spins.

\begin{proposition}[Classical limit]\label{pro:cl-lim}
  For $v\in\ell^1_3(\Lat)$ and $J\in\tfrac12 \N_0$, we have
  \begin{equation*}
    \Bigl| \omega\bigl( e^{\frac iJ \sum_x v_x\cdot S_x}\bigr) - 
    e^{i\sum_x v_x\cdot u_x}
    \Bigr| \leq  \frac1{J} \exp\Bigl[ |v|_1 +  2 \exp\bigl(|v|_1 \bigr) \Bigr]
  \end{equation*}
\end{proposition}

More general results for products of characterisitc functions, analogous to Theorem
\ref{thm:clt-gvv} below, can also be derived for the classical limit, but the above
proposition is sufficient for our needs.

The next step is to study the fluctuations of these observables around their mean and we
define fluctuation observables by
\begin{equation}\label{eq:14}
  F_J(v) = \sqrt{\frac2J} \sum_{x\in\Lat} 
  \bigl[ v_x\cdot S_x - \omega(v_x\cdot S_x)\bigr]
\end{equation}
Our first result here is about the characteristic function of the fluctuation observables
in a coherent state:

\begin{proposition}\label{pro:clt-1}
  For $v\in\ell^2_3(\Lat)$ and  $J\in\tfrac12 \N_0$, we have
  \begin{equation*}
    \Bigl| \omega\bigl( e^{i F_J(v)} \bigr) - 
    e^{-\frac12 |\tilde v|^2_2}\Bigr|\leq
    \frac1{J^{\hlf}} b(v)
  \end{equation*}
  where
  \begin{equation}\label{eq:3}
        b(v) = \exp\Bigl[ 2^{\hlf} |v|_2 +  2^{\hlf} 
        \exp\bigl(    2^{\hlf} |v|_2 \bigr) \Bigr]
  \end{equation}
\end{proposition}

The vector $\tilde v\in\ell^2(\Lat)$ in the proposition whose length determines the
variance of $F_J(v)$ is defined as follows. At each site $x$ we start with the same
standard basis $(e^1_x, e^2_x, e^3_x)$ of $\R^3$. Through the state $\omega$, we are given
a unit vector $u_x=(\theta_x,\phi_x)$ where the spherical coordinates are given w.r.t. the
standard basis. We can define a new basis of $\R^3$ (now site dependent) by rotating
$e^3_x$ to $u_x$. The rotated $e^1_x$ and $e^2_x$ then span the plane orthogonal to $u_x$
which can be identified with the tangent plane to the unit sphere at $u_x$. We define
$\tilde v_x\in\R^2$ to be the projection of $v_x$ onto this plane.

To see that this is really the variance, we can compute directly from eq.  \eqref{eq:2}
that
\begin{equation}\label{eq:7}
    \omega\bigl( F_J(v)F_J(w) \bigr) =\sum_{x\in\Lat}
    \bigl[  v_x\cdot w_x - (v_x\cdot
    u_x)(w_x\cdot   u_x) + i (v_x\times w_x)\cdot u_x \bigr]
\end{equation}
and it follows that indeed
\begin{equation*}
  \omega\bigl( F_J(v)^2\bigr) = |\tilde v|_2^2
\end{equation*}

The proof of Proposition \ref{pro:clt-1} is given in Appendix
\ref{sec:proof-central-limit-intro} as well.

We can continue along the lines of \cites{goderis:1989b, goderis:1990} and show that in
the limit $J\to\infty$ the system of fluctuations is given by a representation of the
canonical commutation relations (\textsf{CCR}). The most general theorem in this context
is:
\begin{theorem}\label{thm:clt-gvv}
  For $n\in\N_0$, $v_1,\dots,v_n\in\ell^2_3(\Lat)$ and $J\in\tfrac12 \N_0$,
  \begin{equation}\label{eq:4}
    \Bigl| \omega\Bigl( \prod_{j=1}^n e^{iF_J(v_j)} \Bigr) 
    - \tilde\omega\Bigl(
    \prod_{j=1}^n W(\tilde v_j) \Bigr)\Bigr|\leq \frac1{J^{\hlf}} 
    \biggl\{ b\Bigl(
    \sum_{j=1}^n v_j\Bigr) + \sum_{j=1}^{n-1} a\Bigl( v_j, 
    \sum_{k=j+1}^n v_k\Bigr) \biggr\}
  \end{equation}
where it is understood that the second sum is zero for $n=1$, $b(v)$ is given
in \eqref{eq:3} and $a(v,w)$ is given by
  \begin{equation*}
    a(v,w) = \frac13 |v|_2 |w|_2 (|v|_2 + |w|_2) + \sqrt{2} \exp\Bigl[ \frac12
    |v|_2 |w|_2 +  \exp\bigl(|v|_2 |w|_2 \bigr) \Bigr]
  \end{equation*}
\end{theorem}

In this theorem, the $\tilde v_j\in\ell^2(\Lat)$ are again defined by projecting onto the
tangent planes at the different $u_x$; the $W(\tilde v)$ are the Weyl operators generating
the \textsf{CCR} algebra $\ccr(\ell^2(\Lat),\sigma)$, and $\sigma$ is the symplectic form
associated with the standard inner product in $\ell^2(\Lat)$, i.e.,
\begin{equation*}
  \sigma(\tilde v, \tilde w) = 2\Im \langle \tilde v, \tilde w\rangle = 2\Im
  \sum_{x\in\Lat} \overline{\tilde v}_x \tilde w_x 
\end{equation*}
and $\tilde \omega$ is the quasi-free Fock state defined by
\begin{equation*}
  \tilde\omega\bigl( W(\tilde v)\bigr) = e^{-\frac12 \langle \tilde v,\tilde v\rangle} =
  e^{-\frac12 |\tilde v|_2^2}
\end{equation*}
The Weyl operators satisfy the well-known canonical commutation relations:
\begin{equation*}
  W(\tilde v) W(\tilde w) = e^{-\frac i2\sigma(\tilde v,\tilde w)} W(\tilde v+\tilde w)
\end{equation*}

The proof of Theorem \ref{thm:clt-gvv} requires a Baker-Campbell-Hausdorff-type formula
(\textsf{BCH} formula) that shows that the $e^{iF_J(v)}$ approximate these commutation
relations in a suitable sense, and then proceeds from Proposition \ref{pro:clt-1} through
a standard induction argument. For convenience of the reader, complete proofs are included
in Appendix \ref{sec:proof-central-limit-intro}.

The proofs of both Proposition \ref{pro:cl-lim} and \ref{pro:clt-1} and Theorem
\ref{thm:clt-gvv} rely directly on the use of the generating function \eqref{eq:2}. In
order to prove convergence of moments or convergence of the characteristic function of
polynomials of the $F_J(v)$, Dyson's spin wave formalism will prove much more convenient.

\section{Dyson's spin wave formalism for coherent states}
\label{sec:dyson}

In his famous papers \cites{dyson:1956,dyson:1956b}, Dyson introduces a formalism for
studying rigorously bosonization in the Heisenberg ferromagnet.  The main idea is to take
the ferromagnetic ground state $\otim_{x\in\Lat} |J\rangle$ as a reference state, and to
identify the lowering of spins in this state with the creation of boson particles in a
Fock state. This allows the Hamiltonian to be written as an operator on Fock space, and it
is argued that the leading terms at low temperature are the ones linear and quadratic in
the creation and annihilation operators, thus obtaining an exactly solvable system.  This
approximation corresponds to taking a large $J$ limit much as we want to do here.  A nice
exposition of Dyson's formalism in a more modern language is in \cite{hemmen:1984}, and a
rigorous theorem about convergence of the free energy within this formalism is in
\cite{conlon:1990}.

Upon inspection it is clear that the same formalism can also be used if the reference
state is a general product state of coherent states instead of the purely ferromagnetic
state $\otim_x |J\rangle$. This was used in \cite{michoel:2003a} to prove that the
low-energy excitations of interface ground states of the $1$-dimensional \textsf{XXZ}
chain are given, in the large $J$ limit, by a quadratic boson Hamiltonian describing
particles hopping on the lattice under the influence of an external potential centred
around the interface. In this paper we use the formalism in an analogous way to prove
several central limit theorems for the fluctuation observables $F_J(v)$.

Recall the new basis we defined for every site $x$ using the unit vectors $u_x$ that
determine the state $\omega$ (see after Proposition \ref{pro:clt-1}).  More precisely,
this rotated basis is given by \label{pg:basis}
\begin{align*}
  f^1_x &= \cos\theta_x\cos\phi_x\;e^1_x + \cos\theta_x\sin\phi_x\;e^2_x -
  \sin\theta_x\;   e^3_x\\
  f^2_x &= -\sin\phi_x\;e^1_x + \cos\phi_x\;e^2_x\\
  f^3_x &= u_x=\sin\theta_x\cos\phi_x\;e^1_x + \sin\theta_x\sin\phi_x\;e^2_x +
  \cos\theta_x\; e^3_x
\end{align*}
The spin operators can be rotated likewise, and we find (using, e.g., eqs.  (3.9) of
\cite{arecchi:1972})
\begin{align*}
  \tilde S^1_{x}&= U_x S^1_{x} U_x^* = f^1_x\cdot S_x =
  \cos\theta_x\cos\phi_x\;S^1_x + \cos\theta_x\sin\phi_x\;S^2_x -
  \sin\theta_x\;  S^3_x\\
  \tilde S^2_x &=U_x S^2_x U_x^* = f^2_x\cdot S_x = -\sin\phi_x\;S^1_x +
  \cos\phi_x\;S^2_x\\
  \tilde S^3_x &=U_x S^3_x U_x^* = f^3_x\cdot S_x =
  u_x=\sin\theta_x\cos\phi_x\;S^1_x + \sin\theta_x\sin\phi_x\;S^2_x +
  \cos\theta_x\;   S^3_x
\end{align*}
where
\begin{equation*}
  U_x=e^{\frac12\theta_x( S^-_x e^{i\phi_x} -   S^+_x e^{-i\phi_x})}
\end{equation*}

The main observation is that in this new basis the state
$\otim_x|(\theta_x,\phi_x)\rangle$ becomes the usual reference state $\otim_x |+J\rangle$
for the spin wave formalism:
\begin{equation*}
  \tilde S^3_x|(\theta_x,\phi_x)\rangle = \tilde S^3_x U_x  |J\rangle =
  U_x S^3_x |J\rangle = J   |(\theta_x,\phi_x)\rangle 
\end{equation*}

The rotated spin raising and lowering operators are
\begin{equation*}
  \tilde S^{\pm}_x = \tilde S^1_x \pm i\tilde S^2_x
\end{equation*}
or
\begin{align*}
  \tilde S^+_x &= -\sin\theta_x S^3_x + 
  \cos^2(\tfrac12 \theta_x)e^{-i\phi_x} S^+_x
  - \sin^2(\tfrac12 \theta_x)e^{i\phi_x} S^-_x\\
  \tilde S^+_x &= -\sin\theta_x S^3_x - 
  \sin^2(\tfrac12 \theta_x)e^{i\phi_x} S^+_x
  + \cos^2(\tfrac12 \theta_x)e^{-i\phi_x} S^-_x
\end{align*}

According to our previous notation we denote the components of a vector $v_x\in\R^3$
w.r.t. the standard basis with $v^i_x$, and w.r.t. the rotated basis with $\tilde v^i_x$.
We find
\begin{equation*}
  v_x\cdot S_x = \frac12 \tilde v^-_x \tilde S^+_x 
  + \frac12 \tilde v^+_x \tilde S^-_x +
  \tilde   v^3_x \tilde S^3_x
\end{equation*}
where $\tilde v^\pm_x = \tilde v^1_x \pm i \tilde v^2_x$, and 
\begin{equation*}
  F_J(v) = \sum_{x\in\Lat} \tilde v^-_x \frac{\tilde S^+_x}{(2J)^{\hlf}} 
  +  \tilde
  v^+_x \frac{\tilde S^-_x}{(2J)^{\hlf}}  
  + \bigl(\frac 2J\bigr)^{\hlf} \tilde v^3_x
  (\tilde S^3_x - J)
\end{equation*}
where we used $\omega(v_x\cdot S_x)=Jv_x\cdot u_x= J\tilde v^3_x$.

Let
\begin{align*}
  \vec n &= \{ n_x\in\N\}_{x\in\Lat}\\
  \mathcal{N}_J&= \Bigl\{ \vec n\mid \forall x\colon n_x\leq 2J ,
  \sum_x n_x <\infty \Bigr\}\\
  \phi_{\vec n} &= \prod_{x\in\Lat} \frac1{n_x!} 
  \binom{2J}{n_x}^{-\frac12} \bigl(\tilde
  S^-_x\bigr)^{n_x} \Omega_J
\end{align*}
where $\Omega_J=\otim_x|(\theta_x,\phi_x)\rangle$ is the \textsf{GNS} vector for the state
$\omega$.  The set $\{\phi_{\vec n}\mid \vec n\in\mathcal{N}_J\}$ is an orthonormal basis
for $\Hi$, which will prove to be very useful.

Recall that $\tilde\omega$ is the quasi-free Fock state on $\ccr(\ell^2(\Lat),\sigma)$.
Its \textsf{GNS} Hilbert space is the usual Fock space $\mathcal{F}$ with a vacuum vector
$\tilde\Omega$, and creation and annihilation operators $a^*_x$ and $a_x$:
\begin{align*}
  [a_x,a^*_y]&=\delta_{xy}\\
  a_x\tilde\Omega &= 0
\end{align*}
The representative of $W(\tilde v)$ is the usual Weyl operator
\begin{equation*}
  W(\tilde v) = e^{iF(\tilde v)} 
  = e^{i\sum_x \tilde v^+_x a^*_x + \tilde v^-_x a_x}
\end{equation*}
where, if $\tilde v_x=(\tilde v^1_x,\tilde v^2_x)\in\R^2$, $\tilde v^\pm_x\in\C$ are
defined as above. The unbounded operator
\begin{equation}\label{eq:5}
  F(\tilde v)=\sum_{x\in\Lat} \bigl[ \tilde v^+_x a^*_x 
  + \tilde v^-_x a_x \bigr]
\end{equation}
is well-defined for $\tilde v\in \ell^2(\Lat)$ and is called the boson field operator.

Now let
\begin{align*}
  \mathcal{N}&= \Bigl\{ \vec n\mid   \sum_x n_x <\infty \Bigr\}\\
  \tilde\phi_{\vec n} &= \prod_{x\in\Lat} \frac1{(n_x)^{\hlf}} (a^*_x)^{n_x}\tilde\Omega
\end{align*}
then we have an orthonormal basis $\{\tilde\phi_{\vec n}\mid \vec n\in\mathcal{N}\}$ of
$\mathcal{F}$. Identifying $\phi_{\vec n}$ with $\tilde\phi_{\vec n}$, it is clear that
the \textsf{GNS} Hilbert spaces, $\Hi$, $J\in\frac12\N_0$, can be identified with a nested
sequence of subspaces of $\mathcal{F}$, defined for each $J$ as the linear span of all
vectors $\tilde\phi_{\vec n}$, with $n_x\leq 2J$.  More precisely, we use the projections
$P_{n,x}$ on $\mathcal{F}$ which project onto the first $2n$ boson states at site $x$,
i.e., on the states $\tilde\phi_{\vec n}$ with $0\leq n_x\leq 2n$, and denote $P_n =
\prod_x P_{n,x}$, and find
\begin{equation*}
  \Hi = P_J \mathcal{F}
\end{equation*}
where $=$ means unitarily equivalent.

Under this equivalence, we find that the spin operators are given by (see
\cite{hemmen:1984} for more details)
\begin{align*}
  \frac{\tilde S^-_x}{(2J)^{\hlf}} = P_J a^*_x g_J(x)^{\hlf}, && 
  \frac{\tilde
    S^+_x}{(2J)^{\hlf}} = g_J(x)^{\hlf} a_x P_J, && J-\tilde S^3_x 
    = P_J a^*_x a_x P_J
\end{align*}
where
\begin{equation*}
  g_J(x) = g_J(a^*_x a_x)
\end{equation*}
and
\begin{equation*}
  g_J(n)=
  \begin{cases}
    1-\frac 1{2J} n & n\leq 2J\\
    0 & n>2J
  \end{cases}
\end{equation*}
Hence
\begin{equation}\label{eq:19}
  F_J(v) = P_J\biggl\{ \sum_{x\in\Lat} \tilde v^+_x a^*_x g_J(x)^{\hlf} 
  + \tilde v^-_x g_J(x)^{\hlf} a_x - \bigl(\frac2J\bigr)^{\hlf} 
  \tilde v^3_x a^*_x a_x\biggr\} P_J
\end{equation}
If we let $J\to\infty$, we have, for fixed $n$, $g_J(n)\to 1$ and formally $F_J(v)\to
F(\tilde v)$, the boson field operator of eq. \eqref{eq:5}. To make this into a
mathematically precise statement is the subject of the next section.

\section{Operator convergence of fluctuation operators}
\label{sec:clt-spin-wave}

For any $n\in\N$, let $\mathcal{D}_n$ denote the linear span of the vectors
\begin{equation*}
  F(\tilde w_1)\dots F(\tilde w_n)\tilde\Omega \colon \tilde w_1,\dots, 
  \tilde w_n\in
  \ell^2(\Lat)
\end{equation*}
and define $\mathcal{D}$ by
\begin{equation*}
  \mathcal{D}=\bigoplus_{n\in\N}\mathcal{D}_n,
\end{equation*}
i.e., $\mathcal{D}$ is the linear span of the vectors $\tilde\phi_{\vec n}$, $\vec
n\in\mathcal{N}$.  The set $\mathcal{D}$ consists of entire analytic vectors for $F(\tilde
v)$, $\tilde v\in\ell^2(\Lat)$ (Theorem 4.6 of \cite{petz:1990}), i.e.,
$\psi\in\mathcal{F}$ such that $\psi$ is in the domain of $F(\tilde v)^k$ for every
$k\in\N$ and
\begin{equation*}
  \sum_{k\geq 0}\frac{t^k}{k!}\bigl\| F(\tilde v)^k \psi\bigr\| 
  <\infty \quad (t>0)
\end{equation*}
and therefore $\mathcal{D}$ is dense in $\mathcal{F}$. It also follows that $\mathcal{D}$
is a core for all $F(\tilde v)$, $\tilde v\in\ell^2(\Lat)$ as well as for the creation and
annihilation operators $a^*(\tilde v)$ and $a(\tilde v)$,
\begin{equation*}
  a^*(\tilde v) = [a(\tilde v)]^* = \sum_{x\in\Lat} \tilde v^+_x a^*_x
\end{equation*}

Clearly, any function of the number operators, $n_x=a^*_x a_x$, such as $P_J$, leaves the
spaces $\mathcal{D}_n$ invariant and hence:
\begin{lemma}\label{lem:F_J-in-D}
  For all $n\in\N$, $v_1,\dots,v_n\in\ell^1_3(\Lat)$, 
  and $J\in\frac12\N_0$ such that
  $2J>n$, we have
  \begin{equation*}
    F_J(v_1)\dots F_J(v_n)\tilde\Omega \in\mathcal{D}_n
  \end{equation*}
\end{lemma}

Recall the following lemma, Lemma 4.5 of \cite{petz:1990}:
\begin{lemma}\label{lem:petz}
  For all $n\in\N$, $\psi_n\in\mathcal{D}_n$ and $\tilde v\in \ell^2(\Lat)$, 
  we have
  \begin{equation*}
    \bigl\| F(\tilde v) \psi_n\bigr\|  \leq 2|\tilde v|_2 (n+1)^{\hlf} 
    \bigl\|\psi_n\bigr\|  
  \end{equation*}
\end{lemma}

The analogous result for the $F_J(v)$ is:
\begin{lemma}\label{lem:petz-J}
  For all $n\in\N$, $\psi_n\in\mathcal{D}_n$, $v\in\ell^1_3(\Lat)$, 
  and $J\in\frac12\N_0$
  such that $2J>n$, we have
  \begin{equation*}
    \bigl\| F_J(v) \psi_n\bigr\| \leq  4 |v|_1 (n+1)^{\hlf}    
    \bigl\|\psi_n\bigr\|  
  \end{equation*}
\end{lemma}
\begin{proof}
  By the choice $2J>n$ we do not need the projection operators $P_J$, in other words
  $\mathcal{D}_n\subset\Hi = P_J\mathcal{F}$.  For simplicity, denote
  \begin{equation*}
    A_x = \tilde v^-_x g_J(x)^{\hlf} a_x
  \end{equation*}
  We have
  \begin{equation*}
    F_J(v) = \sum_x A^*_x + A_x - \bigl(\frac 2J\bigr)^{\hlf} 
    \tilde v^3_x  a^*_x a_x
  \end{equation*}
  and
  \begin{equation*}
    \bigl\| F_J(v)  \psi_n \bigr\| \leq \sum_x \bigl\| (A^*_x +
    A_x)\psi_n \bigr\| + \bigl(\frac 2J\bigr)^{\hlf} 
    \sum_x |\tilde v^3_x| \bigl\| a^*_x a_x
    \psi_n \bigr\|
  \end{equation*}
  The second term is bounded by $2^{\hlf}J^{-\hlf} n |v|_1 \|\psi_n\|$; for the first term
  we use
  \begin{equation*}
    \bigl\| (A^*_x +  A_x)\psi_n \bigr\|^2 \leq 2\bigl\| (A^*_xA_x)^{\hlf} 
    \psi_n\bigr\|^2 +
    2\bigl\| (A_xA^*_x)^{\hlf} \psi_n\bigr\|^2 
  \end{equation*}
  On $\mathcal{D}_n$, we have, using $2J>n$ as well,
  \begin{align*}
    A^*_x A_x &= |\tilde v_x|^2 a^*_x g_J(x) a_x 
    = |\tilde v_x|^2\Bigl[ n_x \bigl(
    1-\frac{n_x}{2J}\bigr) +  \frac{n_x}{2J}\Bigr] 
    \leq |\tilde v_x|^2 (n+1) \I\\
    A_x A^*_x &= |\tilde v_x|^2 g_J(x)^{\hlf} a_xa^*_x g_J(x)^{\hlf} 
    = |\tilde v_x|^2
    (n_x+1) \bigl( 1-\frac{n_x}{2J} \bigr)\leq |\tilde v_x|^2 (n+1) \I
  \end{align*}
  and, using once more $2J>n$,
  \begin{equation*}
    \bigl\|  F_J(v)  \psi_n \bigr\| 
    \leq  2 |v|_1 \|\psi_n\| \Bigl[ (n+1)^{\hlf} +
    \frac n{(2J)^{\hlf}} \Bigr]\leq 4 |v|_1 (n+1)^{\hlf} \|\psi_n\|
  \end{equation*}
\end{proof}

Denote with $\slim$ the strong resolvent operator limit for operators acting on
$\mathcal{F}$. The following is our general result about the convergence of the
fluctuation observables.

\begin{theorem}\label{thm:clt-dyson}
  For all $k\in\N$ and $v_1, \dots, v_k\in\ell^1_3(\Lat)$, we have
  \begin{equation*}
    \slim_{J\to\infty} \prod_{j=1}^k F_J(v_j) = \prod_{j=1}^k F(\tilde v_j)
  \end{equation*}
  where the limit is taken over any sequence of $J\in\frac12\N_0$ tending to $\infty$.
  More precisely, for all $n\in\N$, $\psi_n\in\mathcal{D}_n$, and $J\in\frac12\N_0$ such
  that $2J>n+k$, we have
  \begin{equation}\label{eq:13}
    \Bigl\| \Bigl[ \prod_{j=1}^k F(\tilde v_j) - \prod_{j=1}^k F_J(v_j) \Bigr]  
    \psi_n \Bigr\|
    \leq \frac1{(2J)^{\hlf}} \Bigl[ \prod_{j=1}^{k} |v_j|_1 \Bigr] \Bigl[
    \frac{(n+k)!}{n!}\Bigr]^{\hlf} \sum_{i=1}^k 2^{k+i} (n+k-i)^{\hlf} 
    \bigl\|\psi_n\bigr\|
  \end{equation}
\end{theorem}

It is convenient to single out the $k=1$ case as a separate result, and prove this first,
which we do in the following lemma. The proof of the Theorem \ref{thm:clt-dyson} is then
given after the proof of Lemma \ref{lem:clt-dyson-1}.

\begin{lemma}\label{lem:clt-dyson-1}
  For $v\in\ell^1_3(\Lat)$, we have
  \begin{equation}\label{eq:6}
    \slim_{J\to\infty} F_J(v) = F(\tilde v)
  \end{equation}
  where the limit is taken over any sequence of $J\in\frac12\N_0$ tending to $\infty$.
  More precisely, for all $n\in\N$, $\psi_n\in\mathcal{D}_n$, and $J\in\frac12\N_0$ such
  that $2J>n+1$, we have
  \begin{equation}\label{eq:12}
    \bigl\| \bigl[ F(\tilde v) - F_J(v) \bigr]\psi_n \bigr\| 
    \leq 4 |v|_1 \frac
    n{(2J)^{\hlf}} \bigl\|\psi_n\bigr\| 
  \end{equation}
\end{lemma}
\begin{proof}
  First note that \eqref{eq:6} follows directly from \eqref{eq:12}, see, e.g., Theorem
  \textsf{VIII}.25 of \cite{reed:1972}, so it is sufficient to prove \eqref{eq:12}. This
  can be done as in the proof of Lemma \ref{lem:petz-J}.  This time,
  \begin{equation*}
    A_x = \tilde v^-_x \bigl( 1- g_J(x)^{\hlf} \bigr) a_x
  \end{equation*}
  We have
  \begin{equation*}
    \bigl\| \bigl[ F(\tilde v) - F_J(v) \bigr] \psi_n \bigr\| 
    \leq \sum_x \bigl\| (A^*_x +
    A_x)\psi_n \bigr\| + \bigl(\frac 2J\bigr)^{\hlf} 
    \sum_x |\tilde v^3_x| \bigl\| a^*_x a_x
    \psi_n \bigr\|
  \end{equation*}
  The second term is bounded by $2^{\hlf}J^{-\hlf} n |v|_1 \|\psi_n\|$; for the first term
  we use again
  \begin{equation*}
    \bigl\| (A^*_x +  A_x)\psi_n \bigr\|^2 \leq 2\bigl\| (A^*_xA_x)^{\hlf} 
    \psi_n\bigr\|^2 +
    2\bigl\| (A_xA^*_x)^{\hlf} \psi_n\bigr\|^2 
  \end{equation*}
  These two terms can be bounded using $1-g_J(x)^{\hlf}\leq n_x (2J)^{-1}$ and Lemma
  \ref{lem:petz}:
  \begin{align*}
    \bigl\| (A^*_xA_x)^{\hlf} \psi_n\bigr\|^2 
    &= \langle \psi_n, A^*_x A_x \psi_n\rangle
    = |\tilde v_x|^2 \langle a_x\psi_n, ( 1-g_J(x)^{\hlf})^2 a_x\psi_n\rangle\\
    &\leq |\tilde v_x|^2 \bigl(\frac n{2J}\bigr)^2 
    \bigl\| a_x\psi_n\|^2 \leq |\tilde
    v_x|^2 \bigl(\frac  n{2J}\bigr)^2 n    \|\psi_n\|^2 \\
    \bigl\| (A_xA^*_x)^{\hlf} \psi_n\bigr\|^2 
    &= \langle \psi_n, A_x A^*_x \psi_n\rangle =
    |\tilde v_x|^2\langle \psi_n, ( 1-g_J(x)^{\hlf}) 
    a_x a^*_x ( 1-g_J(x)^{\hlf}) \psi_n\rangle\\
    &\leq |\tilde v_x|^2 \bigl(\frac n{2J}\bigr)^2 (n+1) \|\psi_n\|^2
  \end{align*}
  Putting everything together, and using $2J>n+1$, we find
  \begin{equation*}
    \bigl\| \bigl[ F(\tilde v) - F_J(v) \bigr] \psi_n \bigr\| 
    \leq 2 |v|_1 \Bigl[
    \frac{n(n+1)^{\hlf}}{2J} + \frac n{(2J)^{\hlf}}\Bigr]  \|\psi_n\| 
    \leq 4 |v|_1 \frac n{(2J)^{\hlf}} \|\psi_n\|
  \end{equation*}
\end{proof}

Now follows the remainder of the proof of  Theorem \ref{thm:clt-dyson}.
\begin{proof}[Proof of Theorem \ref{thm:clt-dyson}]
  As in the proof of Lemma \ref{lem:clt-dyson-1}, it is again sufficient to prove eq.
  \eqref{eq:13}. We write the difference of products as a telescopic sum:
  \begin{align*}
    \prod_{j=1}^{k} F(\tilde v_j) - \prod_{j=1}^{k} F_J(v_j)
    &=\bigl[ F(\tilde v_1) - F_J(v_1)\bigr] 
    F(\tilde v_2) \dots F(\tilde v_k)\\
    &\quad+ F_J(v_1) \bigl[ F(\tilde v_2) - F_J(v_2)\bigr] 
    F(\tilde v_3) \dots F(\tilde
    v_k) \\
    &\quad+\dots\\
    &\quad+ F_J(v_1)\dots F_J(v_{k-1)}\bigl[ F(\tilde v_k) - F_J(v_k)\bigr]
  \end{align*}
  We estimate each term separately, first using repeatedly Lemma \ref{lem:petz-J}, then
  using Lem\-ma \ref{lem:clt-dyson-1}, and finally using repeatedly Lemma \ref{lem:petz}:
  \begin{align*}
    &\Bigl\| F_J(v_1)\dots F_J(v_{i-1}) 
    \bigl[ F(\tilde v_i) - F_J(v_i)\bigr] F(\tilde
    v_{i+1})    \dots F(\tilde v_k) \psi_n\Bigr\|\\
    &\quad\leq \Bigl[ \prod_{j=1}^{i-1} 4 |v_j|_1(n+k-j+1)^{\hlf} \Bigr] 
    \Bigl\| \bigl[
    F(\tilde v_i) - F_J(v_i)\bigr] F(\tilde v_{i+1})    \dots 
    F(\tilde v_k) \psi_n\Bigr\|\\
    &\quad\leq \Bigl[ \prod_{j=1}^{i-1} 4 |v_j|_1(n+k-j+1)^{\hlf} \Bigr] 
    4 |v_i|_1
    \frac{(n+k-i)}{(2J)^{\hlf}}     \Bigl\| F(\tilde v_{i+1})    
    \dots F(\tilde v_k) \psi_n\Bigr\|\\
    &\quad\leq \Bigl[ \prod_{j=1}^{i-1} 4 |v_j|_1(n+k-j+1)^{\hlf} \Bigr] 
    4 |v_i|_1
    \frac{(n+k-i)}{(2J)^{\hlf}} \Bigl[ \prod_{j=i+1}^k 2 |\tilde v_j|_2
    (n+k-j+1)^{\hlf}\Bigr]\bigl\|\psi_n\bigr\|\\
    &\quad\leq \Bigl[ \prod_{j=1}^{k} |v_j|_1 \Bigr] 4^i 2^{k-i}
    \frac{(n+k-i)^{\hlf}}{(2J)^{\hlf}} 
    \Bigl[ \prod_{j=1}^{k} (n+k-j+1)^{\hlf} \Bigr]
    \bigl\|\psi_n\bigr\|\\ 
    &\quad = 2^{k+i} \Bigl[ \prod_{j=1}^{k} |v_j|_1 \Bigr] 
    \Bigl[ \frac{(n+k)!}{n!}
    \frac{(n+k-i)}{2J} \Bigr]^{\hlf}\bigl\|\psi_n\bigr\|
  \end{align*}
\end{proof}

Now we return to the main subject of this paper, namely proving central limit theorems for
the coherent state $\omega$.

We get convergence of all moments as a trivial application of the Cauchy-Schwarz
inequality and Theorem \ref{thm:clt-dyson}, more specifically eq. \eqref{eq:13} for the
case $n=0$.

\begin{corollary}[Moments]
  For all $k\in\N$ and $v_1, \dots, v_k\in\ell^1_3(\Lat)$, we have
  \begin{equation*}
    \lim_{J\to\infty} \omega\Bigl(\prod_{j=1}^k F_J(v_j)\Bigr) 
    = \tilde\omega\Bigl(
    \prod_{j=1}^k F_J(\tilde v_j) \Bigr)
  \end{equation*}
  where the limit is taken over any sequence of $J\in\frac12\N_0$. More
  precisely, for all $J\in\frac12\N_0$ 
  \begin{equation*}
    \Bigl| \omega\Bigl(\prod_{j=1}^k F_J(v_j)\Bigr) - \tilde\omega\Bigl(
    \prod_{j=1}^k F_J(\tilde v_j) \Bigr)\Bigr| 
    \leq  \frac{(k!)^{\hlf}}{(2J)^{\hlf}} \Bigl[
    \prod_{j=1}^{k} |v_j|_1 \Bigr]  \sum_{i=1}^k 2^{k+i} (k-i)^{\hlf}
  \end{equation*}
\end{corollary}

It is clear that Theorem \ref{thm:clt-dyson} contains much more information about the
convergence of the fluctuation operators than the convergence of all moments that is
derived from it, or the convergence of characteristic functions that is given in Theorem
\ref{thm:clt-gvv}. In the usual setting of quantum central limit theorems
\cites{goderis:1989b, goderis:1990} a space on which all fluctuation operators as well as
the limiting boson field operator act simultaneaously, does not exist and therefore the
question of strong (or any other) operator convergence does not make sense in these
situations. It is an interesting question, however, to ask whether the existing central
limit theorems in the usual setting for sums of random variables can be strengthened.
Therefore, we end this section with several reformulations of Theorem \ref{thm:clt-dyson}
to obtain statements that do make sense although they have not been proved in general. In
the Section \ref{sec:strong-clt-N} we will give a first example of such a stronger
convergence result.

\begin{corollary}[Spectral measure]
  For any $k\in\N$ and $v_1,\dots, v_k\in\ell^1_3(\Lat)$, the spectral measure in the
  coherent state $\omega$ of $F_J(v_1)\dots F_J(v_k)$ converges to the spectral measure in
  the Fock state $\tilde\omega$ of $F(\tilde v_1) \dots F(\tilde v_k)$, where convergence
  is as functionals on $\mathcal{C}_b(\R)$, the bounded continuous functions on $\R$.
\end{corollary}
\begin{proof}
  It follows from Theorem \ref{thm:clt-dyson} and Theorem \textsf{VII}.20 of
  \cite{reed:1972} that for any $f\in\mathcal{C}_b(\R)$, and any
  $\psi\in\mathcal{F}$ \begin{equation*}
    \lim_{J\to\infty} f\Bigl( \prod_{j=1}^k F_J(v_j)\Bigr) \psi 
    = f\Bigl( \prod_{j=1}^k
    F(\tilde v_j)\Bigr) \psi 
  \end{equation*}
\end{proof}

The following reformulation is probably the most interesting. It generalizes Theorem
\ref{thm:clt-gvv} by proving convergence of the characteristic function of arbitrary
polynomials of fluctuation operators instead of only allowing linear combinations of them.

If $A_1,\dots, A_k$ are selfadjoint elements of a general $\cst$ or von Neumann algebra
$\A$, we denote with $\C\langle A_1,\dots, A_k\rangle$ the ring of noncommutative
polynomials in $A_1,\dots, A_k$. This ring admits a $^*$-involution which conjugates each
coefficient and reverses the order of multiplication in each term. A polynomial
$p\in\C\langle A_1,\dots,A_k\rangle$ is called a selfadjoint polynomial if it is invariant
under this involution.  For example, the anticommutator $A_1A_2 + A_2 A_1$ is a
selfadjoint polynomial, but the monomial $A_1A_2$ is not.

\begin{corollary}[Characteristic function of polynomials]\label{cor:strong-clt}
  For any $k\in\N$, $v_1,\dots, v_k\in\ell^1_3(\Lat)$, and $p\in\C\bigl\langle F_J(v_1),
  \dots, F_J(v_k) \rangle$ a selfadjoint polynomial in $k$ variables, we have
  \begin{equation*}
    \lim_{J\to\infty} \omega\Bigl( e^{i p [F_J(v_1),\dots, F_J(v_k)]} \Bigr) =
    \tilde\omega\Bigl( e^{i p [F(\tilde v_1),\dots, F(\tilde v_k)]} \Bigr)
  \end{equation*}
  where the limit is taken over any sequence $J\in\frac12\N_0$ tending to $\infty$.
\end{corollary}
\begin{proof}
  It follows from Theorem \ref{thm:clt-dyson} that
  \begin{equation*}
    \slim_{J\to\infty} p [F_J(v_1),\dots, F_J(v_k)] 
    = p [F(\tilde v_1),\dots, F(\tilde v_k)]
  \end{equation*}
  and by Trotter's theorem (Theorem \textsf{VIII}.21 of \cite{reed:1972}) also
  \begin{equation*}
    \slim_{J\to\infty} e^{i p [F_J(v_1),\dots, F_J(v_k)]} 
    =  e^{i p [F(\tilde v_1),\dots,
      F(\tilde v_k)]} 
  \end{equation*}
  The result follows from the Cauchy-Schwarz inequality in $\mathcal{F}$.
\end{proof}

\begin{remark}
  In the previous corollary we could further generalize and consider $l$ selfadjoint
  polynomials $p_j$ in $k_j$ variables. It follows by the same argument that
  \begin{equation*}
    \lim_{J\to\infty} \omega\Bigl( \prod_{j=1}^l e^{i p_j [F_J(v_{j,1}),\dots,
      F_J(v_{j,k_j})]} \Bigr) 
      =  \tilde\omega\Bigl( \prod_{j=1}^l e^{i p_j [F(\tilde
      v_{j,1}),\dots,  F(\tilde v_{j,k_j})]}\Bigr)
  \end{equation*}
  In addition all products of polynomials and characteristic functions of polynomials
  (e.g., $p_1\times e^{ip_2}$) can be considered as well.
\end{remark}

\begin{remark}
  We have always taken $v\in\ell^1_3(\Lat)$. This is convenient because then the finite
  $J$ fluctuation operators $F_J(v)$ are bounded operators with
  \begin{equation*}
    \bigl\| F_J(v) \bigr\| \leq 2^{\hlf[3]} |v|_1 J^{\hlf}
  \end{equation*}
  We can ask about more general examples where $F_J(v)$ is only densely defined, and one
  such generalization is the following. Consider in expression \eqref{eq:19} the last term
  \begin{equation*}
    \bigl( \frac2J\bigr)^{\hlf} \sum_{x\in\Lat} \tilde v^3_x a^*_x a_x
  \end{equation*}
  On $\mathcal{D}_n$ this operator is bounded by $2^{\hlf} J^{-\hlf} n\sup_{x\in\Lat}
  |\tilde v^3_x|$. Hence all of our results remain unchanged, except for notationally more
  complicated error bounds, if we consider $v\in\tilde\ell_3(\Lat)$ instead of
  $v\in\ell^1_3(\Lat)$, where $\tilde\ell_3(\Lat)$ is defined as the set of $\R^3$-valued
  sequences on $\Lat$ such that $\tilde v=\{(\tilde v^1_x, \tilde v^2_x)\}_{x\in\Lat} \in
  \ell^1(\Lat)$ and $\tilde v^3 = \{ \tilde v^3_x\}_{x\in\Lat} \in \ell_1^\infty(\Z)$.
\end{remark}

\begin{remark}
  In all of the previous results, we can replace the state $\Omega_J=\tilde\Omega$ by a
  perturbed state $\tilde\Omega_P$, as long as the perturbed state is still in the set
  $\mathcal{D}$ of analytic vectors for $F(\tilde v)$.
\end{remark}

\section{Applications}
\label{sec:applications}

\subsection{Bosonization for quantum spin Hamiltonians}
\label{sec:bosonization}

The main application of Theorem \ref{thm:clt-dyson} is for the large-spin
asymptotics of quantum spin Hamiltonians and their corresponding dynamics.
Consider for instance an interaction between spins in a finite volume $\Lambda$
of the type
\begin{equation}\label{eq:15}
  H_{J,\Lambda} =- \sum_{x,y\in\Lambda} \sum_{i,j=1}^3 h_{ij}(x,y) S^i_x S^i_y -
  J \sum_{x\in\Lambda} \sum_{i=1}^3 g_i(x) S^i_x
\end{equation}
where the interaction functions $h_{ij}$ and $g_i$ satisfy all necessary conditions for
selfadjointness of $H_{J,\Lambda}$, and in addition are bounded and of short enough range,
i.e.,
\begin{align*}
  \sup_{x\in\Lat}\Bigl( \sum_{y\in\Lat} \bigl| h_{ij}(x,y)\bigr| \Bigr) &< \infty\\
  \sup_{x\in\Lat} |g_i(x)|&< \infty
\end{align*}
More general Hamiltonians can easily be treated along the same lines, but this example
already includes the various interesting Heisenberg-type models. The minus sign in front
of the Hamiltonian is there for convenience, we do not make any explicit assumptions on
the sign of the $h_{ij}$ or $g_i$.

Bosonization refers to the idea that the low-energy excitations of a quantum spin system
can be effectively described by a boson approximation to \eqref{eq:15}, describing
non-interacting bosons hopping on the lattice $\Lat$.  To make this into a mathematically
precise statement, the Hamiltonian has to be scaled by $J^{-1}$ and a large-spin limit has
to be taken.  Again there exist various ways of taking this limit. One way consists of
proving that the free energy at low temperatures converges to the free energy of a boson
model, see, e.g. \cites{dyson:1956, dyson:1956b, conlon:1990, hemmen:1984}, and this has
been used mainly to study the isotropic Heisenberg ferromagnet. Another approach was
introduced in \cite{michoel:2003a} in the study of $1$-dimensional anisotropic Heisenberg
ferromagnet, and consists of proving strong operator convergence of the Hamiltonian like
we did for the fluctuation observables in Theorem \ref{thm:clt-dyson}. From there,
convergence of eigenvalues and eigenvectors is derived.

In both approaches it is necessary that there is a unique ground state to be used as a
reference state, and this ground state should be a coherent product state. Both the
isotropic and anisotropic Heisenberg model have a rotational symmetry, and hence
uniqueness of the ground state can only be achieved by adding an external field which
breaks this symmetry. For the isotropic model, there is full rotational symmetry, and the
final results do not depend on which ground state is selected.  For the anisotropic model,
there exist non translation invariant ground states describing interfaces, and the
approximating boson model is different for different ground states, see
\cite{michoel:2003a} for details.

In this section we want to demonstrate another use of Theorem \ref{thm:clt-dyson}, namely
in studying the dynamics of (ground) states under perturbations. The main assumption is
that there exists a ground state $\omega$ of \eqref{eq:15} which is a product state of
coherent states, like we used throughout this paper. For the present application it is not
necessary that it is the unique ground state.

The first step is to write the Hamiltonian \eqref{eq:15} in the new natural basis for the
state $\omega$:
\begin{equation*}
  H_{J,\Lambda} 
  =- \sum_{x,y\in\Lambda} \sum_{i,j=1}^3 \tilde h_{ij}(x,y) \tilde S^i_x \tilde
  S^j_y- J \sum_{x\in\Lambda} \sum_{i=1}^3 \tilde g_i(x) \tilde S^i_x
\end{equation*}
where the $\tilde h_{ij}$ and $\tilde g_i$ are easily obtained using the basis
transformation on page \pageref{pg:basis}.  Instead of taking $i,j=1,2,3$ we can also take
$i,j=-,+,3$.  The coherent state in the new basis is just the `all up' state, and this is
an eigenstate of $H_{J,\Lambda}$ if and only if the only non-zero coefficients are as
follows:
\begin{align*}
  H_{J,\Lambda} &=- \sum_{x,y\in\Lambda} 
  \Bigl[ \tilde h_{-+}(x,y) \tilde S^-_x \tilde
  S^+_y + \tilde h_{+-}(x,y) \tilde S^+_x \tilde S^-_y 
  + \tilde h_{33}(x,y) \tilde S^3_x
  \tilde S^3_y \Bigr] - J \sum_{x\in\Lambda} \tilde g_3(x) \tilde S^3_x \\
  &\quad -\sum_{x,y\in\Lambda} 
  \Bigl[ \tilde h_{13}(x,y) \tilde S^1_x \tilde S^3_y +
  \tilde h_{31}(x,y) \tilde S^3_x \tilde S^1_y 
  + \tilde h_{23}(x,y) \tilde S^2_x \tilde
  S^3_y + \tilde h_{32}(x,y) \tilde S^3_x \tilde S^2_y \Bigr]
\end{align*}
For the terms on the first line, no additional conditions except for those of
selfadjointness and summability are needed, the terms on the second line however can only
be allowed if for all $x\in\Lambda$
\begin{align*}
  \sum_{y\in\Lambda} \tilde h_{i3} (x,y) 
  = \sum_{y\in\Lambda} \tilde h_{3i} (x,y) = 0
  \;\text{ for }\; i=1,2
\end{align*}
Suppose the reference state $\omega$ is a ground state of $H_{J,\Lambda}$. Then, its
energy is given by
\begin{equation*}
  - \sum_{x,y\in\Lambda}  \tilde h_{33}(x,y) - \sum_{x\in\Lambda} \tilde g_3(x)
\end{equation*}
Hence we can renormalize the Hamiltonian such that the ground state energy is $0$, and
obtain
\begin{align*}
  H_{J,\Lambda} &= -\sum_{x,y\in\Lambda} 
  \Bigl[ \tilde h_{-+}(x,y) \tilde S^-_x \tilde
  S^+_y + \tilde h_{+-}(x,y) \tilde S^+_x \tilde S^-_y 
  + \tilde h_{33}(x,y)\bigl( \tilde
  S^3_x  \tilde S^3_y-J^2\bigr) \Bigr]\\
  &\quad- J\sum_{x\in\Lambda} \tilde g_3(x) \bigl( \tilde S^3_x -J\bigr) \\
  &\quad -\sum_{x,y\in\Lambda} 
  \Bigl[ \tilde h_{13}(x,y) \tilde S^1_x \bigl(\tilde
  S^3_y-J\bigr) + \tilde h_{31}(x,y) 
  \bigl(\tilde S^3_x -J\bigr) \tilde S^1_y\\
  &\quad\qquad+ \tilde h_{23}(x,y) \tilde S^2_x \bigl(\tilde S^3_y -J\bigr) 
  + \tilde
  h_{32}(x,y) \bigl( \tilde S^3_x - J\bigr) \tilde S^2_y \Bigr]
\end{align*}
With the Hamiltonian correctly renormalized, we can also write down the \textsf{GNS}
Hamiltonian $H_J$ acting on the \textsf{GNS} Hilbert space $\Hi$, it is the same
expression with all sums extended from $\Lambda$ to the entire lattice $\Lat$.

It is also clear that after scaling with $J^{-1}$, this Hamiltonian is of the form
\begin{equation*}
  \tfrac1J H_J = -\sum_{x,y\in\Lat} p^{(2)}_{xy}\bigl( F_J(f^1_x), F_J(f^2_x), 
  F_J(f^3_x)
  \bigr) + \frac 1J \sum_{x,y\in\Lat} \tilde h_{33}(x,y)\bigl( J^2- \tilde
  S^3_x  \tilde S^3_y\bigr) 
  + \sum_{x\in\Lambda} \tilde g_3(x) \bigl( J-\tilde S^3_x
  \bigr)  
\end{equation*}
where $p^{(2)}_{xy}$ are selfadjoint polynomials in $2$ of the $3$ listed variables, and
$f^i_x$ are the rotated basis vectors defined on page \pageref{pg:basis}.

Using Theorem \ref{thm:clt-dyson}, we get immediately that
\begin{equation*}
  \slim_{J\to\infty} 
  \sum_{x,y\in\Lat} p^{(2)}_{xy}\bigl( F_J(f^1_x), F_J(f^2_x), F_J(f^3_x)
  \bigr) = \sum_{x,y\in\Lambda} 
  \Bigl[ \tilde h_{-+}(x,y) a^*_x a_y + \tilde h_{+-}(x,y)
  a_x a^*_y \Bigr]
\end{equation*}
The remaining terms can be evaluated by writing them on Fock space:
\begin{equation*}
  \sum_{x\in\Lambda} \tilde g_3(x) \bigl(  J- \tilde S^3_x  \bigr)  
  = \sum_{x\in\Lambda}
  \tilde g_3(x) n_x
\end{equation*}
and
\begin{align*}
  \tfrac 1J \sum_{x,y\in\Lat} 
  \tilde h_{33}(x,y)\bigl(J^2- \tilde S^3_x \tilde S^3_y\bigr)
  &= \sum_{x,y\in\Lat} \tilde h_{33}(x,y) 
  \Bigl[ J - \tfrac1J (J-n_x)(J-n_y)\Bigr]\\
  &= \sum_{x,y\in\Lat} \tilde h_{33}(x,y) (n_x+n_y - \tfrac 1J n_x n_y)\\
  &= \sum_{x\in\Lat} E(x) n_x - \sum_{x,y\in\Lat} \tilde h_{33}(x,y) 
  \frac{n_x n_y}J
\end{align*}
where
\begin{equation*}
  E(x) = \sum_{y\in\Lat} \bigl[ \tilde h_{33}(x,y) + \tilde h_{33}(y,x)\bigr]
\end{equation*}
Using $\sum_x n_x = n\I$ on $\mathcal{D}_n$, it is seen that the second term above
converges strongly to $0$.

Hence we conclude:
\begin{equation}\label{eq:17}
  \slim_{J\to\infty} \tfrac 1J H_J 
  = \tilde H \defeq \sum_{x\in\Lat} \bigl[ E(x) + \tilde
  g_3(x)\bigr]   a^*_x a_x - \sum_{x,y\in\Lambda} 
  \bigl[ \tilde h_{-+}(x,y) a^*_x a_y +
  \tilde   h_{+-}(x,y) a_x a^*_y \bigr]
\end{equation}
and this is a rigorous way to express bosonization in a quantum spin model which is of the
type \eqref{eq:15} and has a coherent ground state. As the coefficients $\tilde g_3(x)$,
$\tilde h_{-+}(x,y)$, etc., depend on the rotation of the basis, it is clear from this
derivation that the boson Hamiltonian in general will be different for different ground
states. Notice also that it is the second quantization of an operator $H$ on
$\ell^2(\Lat)$ defined by:
\begin{equation}\label{eq:18}
  (Hv)_x = \bigl[ E(x) + \tilde  g_3(x)\bigr] v_x 
  - \sum_{y\in\Lat} \bigl[ \tilde
  h_{-+}(x,y) + \tilde h_{+-}(y,x) \bigr] v_y
\end{equation}

In \cite{michoel:2003a}, eq. \eqref{eq:17} is used to show that the eigenvalues and
eigenvectors of the boson Hamiltonian give the large-spin asymptotics of the eigenvalues
and eigenvectors of the spin Hamiltonian (for the special case of the anisotropic
ferromagnetic Heisenberg chain).  This is a very practical result because the spectrum of
the spin Hamiltonian is generally very hard, if not impossible, to compute for large
systems, while the spectrum of the boson Hamiltonian, which is quadratic, can be computed
very easily, also for large systems (certainly numerically).

Theorem \ref{thm:clt-dyson} can also be used to obtain a rigorous estimate for another
kind of computation, namely for the dynamics of states, typically eigenstates of $H_J$,
under perturbations of the Hamiltonian. Such a problem is typically studied in a weak
coupling limit, where the strength of the perturbation vanishes as the microscopic time
tends to infinity, see \cite{nachtergaele:2002} for a recent example studying the dynamics
of interfaces.

Clearly, if we perturb the spin Hamiltonian by any selfadjoint polynomial in the
fluctuation operators, we can study the influence of this perturbation using the
bosonization approximation, i.e., taking the large-spin limit. More precisely, let $p$ be
a selfadjoint polynomial in $k$ variables and let $v_1,\dots,v_k\in\ell^1_3(\Lat)$. Denote
\begin{align*}
  P_J = p\bigl[ F_J(v_1),\dots, F_J(v_k)\bigr] && 
  \tilde P=p\bigl[ F(\tilde v_1),\dots, F(\tilde
  v_k)\bigr] 
\end{align*}
Then
\begin{equation*}
  \slim_{J\to\infty} \tfrac 1J H_J + P_J = \tilde H + \tilde P
\end{equation*}
and again by Trotter's theorem (Theorem \textsf{VIII}.21 of \cite{reed:1972}),
for all $\psi\in\mathcal{F}$, 
\begin{equation*}
  \lim_{J\to\infty} e^{it (\frac1J H_J + P_J)}\psi = e^{it (\tilde H+P)}\psi
\end{equation*}
While the spin expression on the l.h.s. can again be very difficult to compute, the boson
expression on the r.h.s. can be solved for relevant cases.

Consider for instance the case where the Hamiltonian is perturbed by a fluctuation of the
spin in a certain direction, i.e., $P_J=F_J(v)$ for some $v\in\ell^1_3(\Lat)$. Such a
perturbation was also considered in \cite{goderis:1991b} in the study of spatial
fluctuations of quantum spin systems and their relation with linear response theory. Using
a Dyson expansion and the \textsf{CCR} algebraic structure we find
\begin{equation*}
  e^{it[\tilde H + F(\tilde v)]} e^{-it \tilde H} 
  = e^{-\frac i2\int_0^t ds \sigma(\tilde
    v,\tilde v^s)} e^{i F(\tilde v^t)}
\end{equation*}
where
\begin{equation*}
  \tilde v^t = \int_0^t ds e^{isH} \tilde v
\end{equation*}
and $H$ is given by \eqref{eq:18}. Hence the boson approximation to the time evolution of
the ground state under the perturbed evolution is given by
\begin{equation*}
  e^{it (\tilde H+P)}\tilde\Omega 
  = e^{it[\tilde H + F(\tilde v)]} e^{-it \tilde H}
  \tilde\Omega = e^{-\frac i2\int_0^t ds \sigma(\tilde
    v,\tilde v^s)} e^{i F(\tilde v^t)}\tilde\Omega
\end{equation*}
Using techniques such as in the previous section and such as in Lemma
\ref{lem:commutator-est} below, explicit estimates which vanish for large values of $J$
can be obtained to compare this state with the true evolved state
\begin{equation*}
  e^{it (\frac1J H_J + P_J)}\Omega_J
\end{equation*}

\subsection{Strong central limit theorem for the large $N$ limit of 
$N$ spin-$\frac12$ particles}\label{sec:strong-clt-N}

In this section, we show that Corollary \ref{cor:strong-clt} gives a fully noncommutative
example of Theorem 1 of \cite{kuperberg:2002b}. In this example, we consider a limit where
the number of spin variables, $N$, tends to infinity, while the magnitude of each spin
remains constant equal to 1/2. We change the notation accordingly.

A quantum probability space consists of a $\cst$ or von Neumann algebra $\A$, and a state
$\omega$ on $\A$. Then $(\A^{\otimes N}, \omega_N=\omega^{\otimes N})$ denotes $N$
independent copies of $(\A,\omega)$, and if $A\in\A$, then
\begin{equation*}
  A_i = \I \otimes\dots\otimes \I \otimes A\otimes \I \otimes\dots\otimes \I
\end{equation*}
with $A$ in the $i^{\text{th}}$ position, denotes the $i^{\text{th}}$ copy of $A$.

The fluctuation operator of $A\in\A$ in this context is defined as
\begin{equation*}
  F_N(A) = \frac1{\sqrt N} \sum_{i=1}^N \bigl[ A_i - \omega(A)\bigr]
\end{equation*}
and we can ask the same convergence questions as we did before. In fact, this simple setup
was the first in which the connection between quantum fluctuation operators and
representations of the \textsf{CCR} was worked out in detail \cite{goderis:1989}. In that
paper, the result analogous to Theorem \ref{thm:clt-gvv} was first proved.

The question whether stronger central limit theorems holds in this setup was first raised
in \cite{kuperberg:2002}, and in \cite{kuperberg:2002b} the following special case was
proved:
\begin{theorem*}[Kuperberg \cite{kuperberg:2002b}]
  Let $(\A,\tau)$ be a quantum probability space with $\tau$ a \emph{tracial} state. For
  all $k\in\N$, $A_1,\dots,A_k$ selfadjoint elements of $\A$, and $p\in\C\langle F_N(A_1),
  \dots ,F_N(A_k)\rangle$ a selfadjoint polynomial in $k$ variables, we have
  \begin{equation*} \lim_{N\to\infty} \tau_N\Bigl( e^{ip[F_N(A_1), \dots ,F_N(A_k)]
    }\Bigr) = \mathbb{E}\Bigl( e^{ip[X(A_1),\dots, X(A_k)]}\Bigr)
  \end{equation*}
  where $X(A_1),\dots, X(A_k)$ are \emph{classical} Gaussian random variables with
  covariance matrix \begin{equation*} \mathbb{E}\bigl( X(A_i) X(A_j) \bigr) = \tau(A_iA_j)
  \end{equation*}
\end{theorem*}

The fact that a tracial state leads to classical fluctuations in the limit is to be
expected, it follows directly from the cyclicity of the trace. As far as we know, no
generalization of this theorem to non-tracial states exists.

Consider the simple case in which $\A=\mathbb{M}_2(\C)$, the complex $(2\times
2)$-matrices. $\A$ carries the irreducible spin-$\frac12$ representation of $SU(2)$ given
by $\frac12$ times the Pauli matrices
\begin{align*}
  \sigma^1 = 
  \begin{pmatrix}
    0&1\\ 1&0
  \end{pmatrix}
  && \sigma^2 = 
  \begin{pmatrix}
    0&-i\\ i&0
  \end{pmatrix}
  && \sigma^3 = 
  \begin{pmatrix}
    1&0\\ 0&-1
  \end{pmatrix}
\end{align*}
The standard basis of $\C^2$ is diagonal for $\sigma^3$, we denote $|1\rangle =
\binom{1}{0}$ and $|-1\rangle = \binom{0}{1}$. For the state we take
$\omega=\omega_{\theta,\phi}$, the expectation in the spin-$\frac12$ coherent state
\begin{equation*}
  |(\theta,\phi)\rangle = e^{\frac14 \theta( \sigma^- e^{i\phi} 
  - \sigma^+ e^{-i\phi})}
  |1\rangle
\end{equation*}
where $\sigma^\pm = \sigma^1\pm i\sigma^2$.

We are interested in the selfadjoint operators
\begin{equation*}
  v\cdot\sigma = v^1\sigma^1 + v^2\sigma^2 + v^3\sigma^3,\quad v\in\R^3
\end{equation*}
and their fluctuation operators $F_N(v) \defeq F_N(v\cdot\sigma)$.

But a collection of $N$ independent spin-$\frac12$ degrees of freedom can also be regarded
as one spin-$\frac N2$ degree of freedom, i.e., define
\begin{equation*}
  S^j_N = \sum_{i=1}^N \tfrac12 \sigma^j_i
\end{equation*}
then these operators satisfy the $SU(2)$ commutators as well, and the eigenvalues of each
$S^j$ obviously are $-\frac N2, -\frac N2 +1,\dots, \frac N2$. Moreover we have in the
coherent state $\omega$:
\begin{align*}
  \omega_N\bigl( e^{iv\cdot S}\bigr) 
  &= \omega_N\bigl( e^{\frac i2\sum_i v\cdot
    \sigma_i}\bigr) =   \prod_{i=1}^N 
    \omega\bigl( e^{\frac i2v\cdot\sigma}\bigr)\\
  &= \Bigl\{ \cos\bigl(\tfrac12 |v|\bigr) 
  + i \frac{v\cdot u}{|v|} \cos\bigl(\tfrac12
  |v|\bigr) \Bigr\}^N
\end{align*}
where $u=(\theta,\phi)$. This is the correct generating function for the spin-$\frac N2$
coherent state, cfr. eq.  \eqref{eq:2}. Hence the present setup is exactly equivalent with
the setup in the previous sections if we take the lattice $\Lat$ consisting of a single
point, and check that the normalization of the fluctuation operators is consistent with
eq. \eqref{eq:14}:
\begin{equation*}
  \sqrt{\frac 2{\tfrac12 N}} \bigl[ v\cdot S_N - \omega_N(v\cdot S_N)\bigr]
  = \frac 2{\sqrt N} \sum_{i=1}^N\bigl[ \tfrac12 v\cdot\sigma_i 
  - \tfrac12 \omega(v\cdot
  \sigma_i) \bigr] = F_N(v)
\end{equation*}

It follows that Corollary \ref{cor:strong-clt} remains valid in the present setup, and we
get immediately the follwoing theorem.
\begin{theorem}
  Consider the quantum probability space $(\mathbb{M}_2(\C), \omega)$ where $\omega$ is
  the coherent state defined by the unit vector $u=(\theta,\phi)\in\R^3$.  For all
  $k\in\N$, $v_1,\dots,v_k\in\R^3$, and $p\in\C\langle F_N(v_1), \dots ,F_N(v_k)\rangle$ a
  selfadjoint polynomial in $k$ variables, we have
  \begin{equation*}
    \lim_{N\to\infty} \omega_N\Bigl( e^{ip[F_N(v_1),  \dots ,F_N(v_k)] }\Bigr) 
    =\tilde\omega\Bigl( e^{ip[ F(\tilde v_1),\dots, F(\tilde v_k)]}\Bigr)
  \end{equation*}
  where $\tilde\omega$ is the Fock state on $\mathsf{CCR}(\R^2,\sigma)$, $\sigma$ is the
  standard symplectic form, and $F(\tilde v)$ are the boson field operators for $\tilde
  v\in\R^2$ which is obtained from $v\in\R^3$ by projecting onto the tangent plane to the
  unit sphere at $u$.
\end{theorem}

Note that the limiting object in the above \textsf{CLT}, the Fock state on
$\mathsf{CCR}(\R^2,\sigma)$, is essentially the quantum harmonic oscillator. It
can be regarded as a non-commuting pair (``position'' and ``momentum'') of
Gaussian random variables.

\appendix

\section{Proofs for Section \ref{sec:clt-gvv}}
\label{sec:proof-central-limit-intro}

\begin{proof}[Proof of Proposition \ref{pro:clt-1}]
  The characteristic function of $F_J(v)$ is, using eq. \eqref{eq:2} and some
  straightforward manipulation,
  \begin{align*}
    \omega\bigl( e^{itF_J(v)}\bigr) 
    &= \prod_x e^{-it\sqrt{\frac2J}\omega(v_x\cdot
      S_x)}\omega\bigl( e^{it\sqrt{\frac2J}  v_x\cdot S_x} \bigr)\\
    &= \prod_x e^{-it (2J)^{\hlf} v_x\cdot u_x}
    \Bigl\{ \cos(\frac{t|v_x|}{(2J)^{\hlf}})+ i
    \frac{v_x\cdot u_x}{|v_x|}\sin (\frac{t|v_x|}{(2J)^{\hlf}}) \Bigr\}^{2J}\\
    &= \prod_x \biggl\{ \frac12\Bigl( 1+\frac{v_x\cdot u_x}{|v_x|} \Bigr) 
    \exp\Bigl[
    \frac{it|v_x|}{(2J)^{\hlf}} \Bigl( 1-\frac{v_x\cdot u_x}{|v_x|} \Bigr) 
    \Bigr]\\
    &\quad\qquad+ \frac12\Bigl( 1-\frac{v_x\cdot u_x}{|v_x|} \Bigr) 
    \exp\Bigl[
    \frac{it|v_x|}{(2J)^{\hlf}} \Bigl( 1+\frac{v_x\cdot u_x}{|v_x|} \Bigr)
    \Bigr]\biggr\}^{2J}
  \end{align*}
  From the formula
  \begin{equation*}
    \omega\bigl(e^{itF_J(v)}\bigr) = \exp\Bigl\{\sum_{k=1}^\infty
    \frac{(it)^k}{ k!}\omega_T\bigl(\underbr{k}{F_J(v),\dots,F_J(v)}\bigr) 
    \Bigr\}
  \end{equation*}
  it follows that the $k$-point truncated correlation function is given by
  \begin{equation*}
    \omega_T\bigl(\underbr{k}{F_J(v),\dots,F_J(v)}\bigr) =
    \Bigl({d\over idt}\Bigr)^k\ln \omega\bigl(e^{itF_J(v)}\bigr)  \Bigr|_{t=0}
    = 2J \sum_x \Bigl({d\over idt}\Bigr)^k\ln f_x(t)\Bigr|_{t=0}
  \end{equation*}
  where $f_x(t)$ is the part between $\{\dots\}$ above. Clearly $f_x(0)=1$, $f_x'(0)=0$,
  and hence
  \begin{align*}
    (\ln f_x)'(0) &= (f_x'/f_x)(0) = 0\\
    (\ln f_x)''(0) &= (f_x''/f_x - (f_x')^2/f_x^2)(0) = f_x''(0) \\
    &\vdots\\
    (\ln f_x)^{(k)}(0) &= f_x^{(k)}(0)
  \end{align*}
  We find for $k\geq 2$:
  \begin{align*}
    &\omega_T\bigl(\underbr{k}{F_J(v),\dots,F_J(v)}\bigr) \\
    &= \frac{2J}{i^k} \sum_x \biggl\{ \frac12 \Bigl( 1+\frac{v_x\cdot u_x}{|v_x|} \Bigr)
    \Bigl[ \frac{i|v_x|}{(2J)^{\hlf}} \Bigl( 1-\frac{v_x\cdot u_x}{|v_x|} \Bigr)
    \Bigr]^k\\
    &\quad\qquad+ \frac12 \Bigl( 1-\frac{v_x\cdot u_x}{|v_x|} \Bigr) \Bigl[
    -\frac{i|v_x|}{(2J)^{\hlf}} \Bigl( 1+\frac{v_x\cdot u_x}{|v_x|} \Bigr) \Bigr]^k
    \biggr\}\\
    &= \frac1{2(2J)^{\hlf[k]-1}} \sum_x |v_x|^k \Bigl( 1-\frac{(v_x\cdot u_x)^2}{|v_x|^2}
    \Bigr) \biggl\{ \Bigl( 1-\frac{v_x\cdot u_x}{|v_x|} \Bigr)^{k-1} + (-1)^k \Bigl(
    1+\frac{v_x\cdot u_x}{|v_x|} \Bigr)^{k-1} \biggr\}
  \end{align*}
  For $k=2$ this reduces to eq. \eqref{eq:7}, i.e., $\omega_T( F_J(v),F_J(v))= |\tilde
  v|_2^2$, while for $k\geq 2$, we get an upper bound
  \begin{equation}\label{eq:11}
    \Bigl| \omega_T\bigl(\underbr{k}{F_J(v),\dots,F_J(v)}\bigr) \Bigr| 
    \leq \frac1{(2J)^{\hlf[k]-1}} \sum_x |v_x|^k 
    \Bigl( 1+\frac{|v_x\cdot u_x|}{|v_x|}
    \Bigr)^{k-1} \leq \frac{2^{\hlf[k]} |v|^k_k}{J^{\hlf[k]-1}}
    \leq \frac{2^{\hlf[k]}
      |v|_2^k}{J^{\hlf[k]-1}} 
  \end{equation}
  such that
  \begin{align*}
    \Bigl| \sum_{k=3}^\infty \frac{i^k}{k!}
    \omega_T\bigl(\underbr{k}{F_J(v),\dots,F_J(v)}\bigr) \Bigr|
    &\leq \sum_{k=3}^\infty\frac{2^{\hlf[k]} |v|_2^k}{k!J^{\hlf[k]-1}} 
    \leq \frac
    1{J^{\hlf}} \exp\bigl( 2^{\hlf} |v|_2\bigr)
  \end{align*}
  We have
  \begin{align*}
    &\Bigl| \omega\bigl( e^{iF_J(v)} \bigr) 
    - \tilde\omega\bigl( W(\tilde v)\bigr)
    \Bigr|\\
    &\leq \Bigl| \exp\Bigl\{ -\frac12 |\tilde v|_2 +\sum_{k=3}^\infty 
    \frac{i^k}{
      k!}\omega_T\bigl(\underbr{k}{F_J(v),\dots,F_J(v)}\bigr) \Bigr\} 
      - \exp\Bigl\{
    -\frac12 |\tilde v|_2\Bigr\} \Bigr|\\
    &\leq \Bigl| \exp\Bigl\{ \sum_{k=3}^\infty \frac{i^k}{
      k!}\omega_T\bigl(\underbr{k}{F_J(v),\dots,F_J(v)}\bigr) \Bigr\} - 1\Bigr|
  \end{align*}
  For $z\in\C$
  \begin{equation*}
    |e^z-1| = \bigl| \int_0^1ds\; e^{sz} z\bigr|\leq e^{|z|} |z|
  \end{equation*}
  and hence
  \begin{equation*}
    \Bigl| \omega\bigl( e^{iF_J(v)} \bigr) 
    - \tilde\omega\bigl( W(\tilde v)\bigr)\Bigr| 
    \leq \frac1{J^{\hlf}} \exp\Bigl[ 2^{\hlf} |v|_2 +  2^{\hlf} \exp\bigl(
    2^{\hlf} |v|_2 \bigr) \Bigr]
  \end{equation*}
  where we used $J\geq \frac12$ in the exponent on the r.h.s.
\end{proof}

\begin{proof}[Proof of Proposition \ref{pro:cl-lim}]
  The proof follows from the previous proof. Consider the centred observable
  \begin{equation*}
    M_J(v) = \frac 1J  \sum_x\bigl( v_x\cdot S_x - J(v_x\cdot u_x) \bigr) =
    \frac1{(2J)^{\hlf}} F_J(v)
  \end{equation*}
  then
  \begin{equation*}
    \omega_T\bigl(\underbr{k}{M_J(v),\dots, M_J(v)}\bigr)
    =\frac1{(2J)^{\hlf[k]}} 
    \omega_T\bigl(\underbr{k}{F_J(v),\dots,  F_J(v)}\bigr) 
   \end{equation*}
   and from eq. \eqref{eq:11}, for $k\geq 2$:
   \begin{equation*}
     \Bigl|\omega_T\bigl(\underbr{k}{M_J(v),\dots, M_J(v)}\bigr)\Bigr| 
     \leq \frac{
       |v|_1^k}{J^{k-1}}  
   \end{equation*}
   Continuing as in the proof of Proposition \ref{pro:clt-1} we find
   \begin{equation*}
      \Bigl| \omega\bigl( e^{\frac iJ \sum_x v_x\cdot S_x}\bigr) 
      - e^{i\sum_x v_x\cdot u_x}
      \Bigr| \leq \Bigl| \omega\bigl( e^{iM_J(v)}\bigr) - 1\Bigr|
      \leq \frac1{J} \exp\Bigl[ |v|_1 +  2 \exp\bigl(|v|_1 \bigr) \Bigr]
   \end{equation*}
\end{proof}

The first step in the proof of Theorem \ref{thm:clt-gvv} is the following Lemma.
\begin{lemma}[A \textsf{BCH}-formula]\label{lem:clt-gvv-bch}
  Let $v_1,v_2\in\ell^2_3(\Lat)$, then, for $J\in\tfrac12 \N_0$,
  \begin{align*}
    \Bigl\| e^{iF_J(v_1)} e^{iF_J(v_2)} - e^{i F_J(v_1+v_2)} e^{-\frac12
      [F_J(v_1),F_J(v_2)]} \Bigr\| \leq \frac1{3 J^{\hlf}} |v_1|_2 |v_2|_2 
      \bigl( |v_1|_2
    + |v_2|_2 \bigr)
  \end{align*}
  where $\|\cdot\|$ is the operator norm in $\mathcal{B}(\Hi)$.
\end{lemma}

This Lemma is a special case of 
\begin{lemma}\label{lem:commutator-est}
  Let $\A$ be a \cst-algebra, $A,B\in\A$ selfadjoint, and $C\in\A$ arbitrary, then
  \begin{align}
    \bigl\|[ e^{iA},C]\bigr\| &\leq  \|[A,C]\| \label{eq:8}\\
    \Bigl\| e^{iA}e^{iB} - e^{i(A+B)}e^{-{1\over 2}[A,B]} \Bigr\|
    &\leq {1\over 3}\Bigl( \bigl\| [A,[A,B]] \bigr\| + \bigl\|
    [B,[B,A]] \bigr\| \Bigr) \label{eq:9}
  \end{align}
\end{lemma}
\begin{proof}
  Eq. \eqref{eq:8} is well-known and proved as follows:
  \begin{align}
    e^{isA}Ce^{-isA}&=C+i\int_0^sdt\;e^{i(s-t)A}[A,C]e^{-i(s-t)A}
    \nonumber \\
    \intertext{and hence}
    [e^{isA},C]&=i\int_0^sdt\;e^{i(s-t)A}[A,C]e^{itA}\label{eq:10}
    \intertext{Therefore}
    \bigl\|[ e^{iA},C]\bigr\|
    &\leq \int_0^1dt\; \bigl\|e^{i(1-t)A}[A,C]e^{itA}\bigr\|
    \leq \| [A,C] \| \nonumber
  \end{align}
  Next we prove \eqref{eq:9}.
  \begin{align*}
    \Bigl\| e^{iA}e^{iB} - e^{i(A+B)}e^{-{1\over 2}[A,B]} \Bigr\| 
    &=\Bigl\| e^{iA}e^{iB}\bigl(1-e^{-iB}e^{-iA} e^{i(A+B)}e^{-{1\over
        2}[A,B]}\bigr) \Bigr\| \\
    &\leq \Bigl\| 1- e^{-iB}e^{-iA} e^{i(A+B)}e^{-{1\over 2}[A,B]} \Bigr\|
  \end{align*}
  Let
  \begin{equation*}
    F(t) = 1- e^{-iB}e^{-itA} e^{i(tA+B)}e^{-{t\over 2}[A,B]}
  \end{equation*}
  We compute $F'(t)$: we have
  \begin{align}
    {d\over dt}e^{i(tA+B)} &= i\int_0^1ds\; e^{is(tA+B)} A e^{i(1-s)(tA+B)} \label{eq:16}
  \end{align}
  and hence
  \begin{align*}
    F'(t)&= i e^{-iB}e^{-itA} A e^{i(tA+B)}e^{-{t\over 2}[A,B]} - i\int_0^1ds\;
    e^{-iB}e^{-itA}e^{is(tA+B)}  A e^{i(1-s)(tA+B)}e^{-{t\over 2}[A,B]}  \\
    &\quad + {1\over 2} e^{-iB}e^{-itA}e^{i(tA+B)} [A,B] e^{-{t\over 2}[A,B]}
  \end{align*}
  Compute, using \eqref{eq:10},
  \begin{align*}
    &iA e^{i(tA+B)}-i\int_0^1ds\; e^{is(tA+B)}A e^{i(1-s)(tA+B)}\\
    &\qquad= i\int_0^1ds\; \bigl( A e^{is(tA+B)} e^{i(1-s)(tA+B)}
    -e^{is(tA+B)}A e^{i(1-s)(tA+B)}\bigr) \\
    &\qquad= i\int_0^1 ds\; [A,e^{is(tA+B)}]e^{i(1-s)(tA+B)}\\
    &\qquad=-\int_0^1ds\int_0^sdu\;e^{i(s-u)(tA+B)}[A,B]e^{i(1-s+u)(tA+B)}
  \end{align*}
  and also
  \begin{equation*}
    \frac12 e^{i(tA+B)}[A,B] =\int_0^1ds\int_0^sdu\;  e^{i(s-u)(tA+B)}
    e^{i(1-s+u)(tA+B)}[A,B]  
  \end{equation*}
  Hence
  \begin{equation*}
    F'(t)= \int_0^1ds\int_0^sdu\; e^{-iB}e^{-itA}
    e^{i(s-u)(tA+B)} \bigl[ e^{i(1-s+u)(tA+B)}, [A,B]\bigr]
    e^{-{t\over 2}[A,B]}  
  \end{equation*}
  We have
  \begin{equation*}
    F(1) = F(0) + \int_0^1dt\;F'(t)
  \end{equation*}
  and hence, using \eqref{eq:8},
  \begin{align*}
    \| F(1)\| &\leq \int_0^1dt\int_0^1ds\int_0^sdu\;
    \Bigl\| \bigl[ e^{i(1-s+u)(tA+B)}, [A,B]\bigr] \Bigr\|\\
    &\leq \int_0^1dt\int_0^1ds\int_0^sdu\; |1-s+u| \bigl\|
    [tA+B,[A,B]] \bigr\|\\
    &\leq \int_0^1ds\int_0^sdu\; |1-s+u| \Bigl( \bigl\|[A,[A,B]]\bigr\| +
    \bigl\|[B,[A,B]]\bigr\|\Bigr)
  \end{align*}
  we find
  \begin{equation*}
    \Bigl\| e^{iA}e^{iB} - e^{i(A+B)}e^{-{1\over 2}[A,B]} \Bigr\| 
    \leq {1\over 3}\Bigl(
    \bigl\|[A,[A,B]]\bigr\| + \bigl\|[B,[B,A]]\bigr\|\Bigr)
  \end{equation*}
\end{proof}

\begin{proof}[Proof of Lemma \ref{lem:clt-gvv-bch}]
  First note that by the $SU(2)$ commutation relations
  \begin{align*}
    \bigl[ F_J(v), F_J(w) \bigr] = \frac1J \sum_x [v_x\cdot S_x, w_x\cdot S_x]
    = \frac1J \sum_x (v_x\times w_x)\cdot S_x
  \end{align*}
  and hence by the previous lemma
  \begin{align*}
    &\Bigl\| e^{iF_J(v)} e^{iF_J(w)} 
    - e^{i F_J(v+w)} e^{-\frac12 [F_J(v),F_J(w)]} \Bigr\|
    \\
    &\quad \leq \frac 1{3J^{\hlf[3]}} \sum_x \Bigl\| 
    \bigl( v_x\times (v_x\times w_x) +
    w_x\times (w_x\times v_x) \bigr)\cdot S_x\Bigr\|\\
    &\quad \leq \frac1{3 J^{\hlf}} 
    \sum_x \bigl| v_x\times (v_x\times w_x) + w_x\times
    (w_x\times v_x) \bigr|\\
    &\quad\leq \frac1{3 J^{\hlf}} \sum_x |v_x| |w_x| (|v_x|+|w_x|) 
    \leq \frac1{3 J^{\hlf}}
    |v|_2 |w|_2 \bigl( |v|_2 + |w|_2 \bigr)
  \end{align*}
\end{proof}

\begin{proof}[Proof of Theorem \ref{thm:clt-gvv}]
  It remains to prove the induction step. Assume \eqref{eq:4} holds for some $n\in\N_0$,
  and choose $v_1,\dots, v_{n+1}\in\ell^2_3(\Lat)$.  For convenience, denote
  \begin{equation*}
    W_J = e^{iF_J(v_1)}\dots e^{iF_J(v_{n-1})}
  \end{equation*}
  By the \textsf{BCH}-formula
  \begin{multline*}
    \Bigl|\omega\Bigl(W_J\bigl[ e^{iF_J(v_n)} e^{iF_J(v_{n+1})} 
    - e^{iF_J(v_n+v_{n+1})}
    e^{-{1\over 2}[F_J(v_n),  F_J(v_{n+1})]}\bigr] \Bigr) \Bigr| \\
    \leq \frac 1{3J^{\hlf}} |v_n|_2 |v_{n+1}|_2 (|v_n|_2 + |v_{n+1}|_2)
  \end{multline*}
  Using the Cauchy-Schwarz inequality, and Proposition \ref{pro:cl-lim},
  \begin{align*}
    &\Bigl| \omega\Bigl( W_J e^{iF_J(v_n+v_{n+1})} 
    \bigl[ e^{-{1\over 2}[F_J(v_n),
      F_J(v_{n+1})]} - e^{-{i\over 2}
      \sigma(v_n,v_{n+1})} \bigr] \Bigr)\Bigr|^2\\
    &\leq 2 - \omega\Bigl(e^{{1\over 2}[F_J(v_n), F_J(v_{n+1})]} \Bigr) 
    e^{-{i\over 2}
      \sigma(v_n,v_{n+1})} - \omega\Bigl(e^{-{1\over 2}[F_J(v_n),
      F_J(v_{n+1})]}
    \Bigr) e^{{i\over 2} \sigma(v_n,v_{n+1})}\\
    &\leq \frac 2J \exp\Bigl[ |v_n|_2 |v_{n+1}|_2 
    + 2 \exp\bigl(|v_n|_2|v_{n+1}|_2 \bigr)
    \Bigr]
  \end{align*}
  Using these two inequalities, and the \textsf{CCR}-algebraic structure, we find
  \begin{align*}
    &\biggl| \omega\Bigl(W_J e^{iF_J(v_n)} e^{iF_J(v_{n+1})} \Bigr) -
    \tilde\omega\Bigl( W(v_1)\dots W(v_n)\Bigr) \biggr| \\
    &\leq \biggl| \omega\Bigl(W_J e^{iF_J(v_n)} e^{iF_J(v_{n+1})} \Bigr) 
    - \omega\Bigl(W_J
    e^{iF_J(v_n+v_{n+1})} e^{-{1\over 2}[F_J(v_n),
      F_J(v_{n+1})]}\bigr] \Bigr) \biggr|\\
    & + \biggl| \omega\Bigl(W_J e^{iF_J(v_n+v_{n+1})} e^{-{1\over 2}[F_J(v_n),
      F_J(v_{n+1})]}\bigr] \Bigr) - \omega\Bigl( W_J e^{iF_J(v_n+v_{n+1})}\Bigr)
    e^{{i\over 2} \sigma(v_n,v_{n+1})}
    \biggr| \\
    &+ \biggl| \omega\Bigl( W_J e^{iF_J(v_n+v_{n+1})}\Bigr) e^{{i\over 2}
      \sigma(v_n,v_{n+1})} - \tilde\omega\Bigl( W(v_1)\dots W(v_{n-1})
    W(v_n+v_{n+1})\Bigr) e^{-{i\over 2}
      \sigma(v_n,v_{n+1})} \biggr| \\
    &+ \biggl| \tilde\omega\Bigl( W(v_1)\dots W(v_{n-1}) 
    W(v_n+v_{n+1})\Bigr) e^{-{i\over
        2} \sigma(v_n,v_{n+1})} - \tilde\omega\Bigl( W(v_1)\dots
    W(v_n)\Bigr)  \biggr|\\
    &\leq \biggl| \omega\Bigl( W_J 
    e^{iF_J(v_n+v_{n+1})}\Bigr) - \tilde\omega\Bigl(
    W(v_1)\dots W(v_{n-1}) W(v_n+v_{n+1})\Bigr)
    \biggr| \\
    &+ \frac 1{J^{\hlf}}
    \biggl\{ \frac13 |v_n|_2 |v_{n+1}|_2 (|v_n|_2 + |v_{n+1}|_2) +
    \sqrt{2} \exp\Bigl[ \frac12 |v_n|_2 |v_{n+1}|_2 
    + \exp\bigl(|v_n|_2|v_{n+1}|_2 \bigr)
    \Bigr] \biggr\}
  \end{align*}
  In the second term, the part between $\{\dots\}$ is the function $a(v_n,v_{n+1})$ as
  defined in the statement of the theorem. Now use the induction hypothesis on the first
  term, and the desired result follows.
\end{proof}

\setlength{\parindent}{0pt}
\setlength{\parskip}{5pt}

\end{document}